\documentclass{article}

\usepackage{graphics}
\usepackage{epsfig}

\begin{document}

\title{Discrete Opinion models as a limit case of the CODA model
}
\author{Andr\'e C. R. Martins\\
GRIFE - EACH, Universidade de S\~ao Paulo, Brazil}

\maketitle

\begin{abstract}
Opinion Dynamics models can be, for most of them, divided between discrete and continuous. They are used in different circumstances and the relationship between them is not clear. Here we will explore the relationship between a model where choices are discrete but opinions are a continuous function (the Continuous Opinions and Discrete Actions, CODA, model) and traditional discrete models. I will show that, when CODA is altered to include reasoning about the influence one agent can have on its own neighbors, agreement and disagreement no longer have the same importance. The limit when an agent considers itself to be more and more influential will be studied and we will see that one recovers discrete dynamics, like those of the Voter model in that limit
%\keywords{Opinion Dynamics; Sociophysics; Bayesian framework; CODA model.}
\end{abstract}

%\ccode{PACS Nos.: 87.23.Ge, 05.65.+b,89.65.-s}

%\pacs{87.23.Ge}{Dynamics of social systems}
%\pacs{05.65.+b}{Self-organized systems}
%\pacs{89.65.-s}{Social and economic systems}

\section{Introduction}

Opinion Dynamics~\cite{castellanoetal07,galametal82,galammoscovici91,sznajd00,stauffer03a,deffuantetal00,hegselmannkrause02} modeling lacks a clear theoretical basis and connections between different models. One unifying proposal exists for discrete opinion models \cite{galam05b}, but it does not include continuous opinions. 
 Understanding how continuous models relate to the discrete ones, if at all, can help us move towards a better understanding of how to describe real social systems. 

Here, I present a variation of the Continuous Opinions and Discrete Actions (CODA) model~\cite{martins08a,martins08b} where an agent considers his own influence in its neighbors. The purpose is both to present the model and to discuss how more traditional models relate to the framework and how they can be seen as approximations or limit cases. In Section \ref{sec:coda}, the relation between discrete spin models, where no probability or strength of opinion exists, and the proposed framework, is explained and I demonstrate for the first time how discrete models can be understood as a limit case of this framework, when agents consider their own influence on their neighbors. Using that demonstration as basis, the original model where each agent considers his own influence on others is introduced in Section \ref{sec:selfinfluence} and we find out that, in the limit of very strong influence, spin dynamics are recovered. We will understand how the model allows any finite system to reach consensus and see that, for finite systems of any size, there is always a range of parameters where the model presented here is identical in results to a spin system. We will also see that, outside the limit, the model has interesting properties about the amount of extremism in the system.

%\section{Traditional existing models and Bayesian version}\label{sec:traditional}

\section{Discrete Opinions and the CODA Model}\label{sec:coda}

In the Continuous Opinions and Discrete Actions (CODA) model~\cite{martins08a,martins08b}, each agent $i$  is trying to decide between two conflicting options. That is, $x$ is a discrete variable with only two possible values, assumed here to be $\pm 1$. This means that the subjective opinion $f_i (x)$ can be trivially described as  $f_i (+1)=p_i$ and, therefore, $f_i (-1)=1-p_i$ . The communication between agents only involve stating which choice is preferred by the agent. That is, what is observed is as spin $s_i$, given by $s_i=A_i[f]=sign(p_i-0.5)$. Finally, the likelihood can be chosen in the simpler possible way, that is, each agent considers there is a chance $p(s_j=+1 | x=+1)=a>0.5$. That is, everyone assigns the same fixed chance $a$ greater than 50\% that a neighbor will choose the best alternative.

With the introduction of a social network that specifies who can be influenced by whom, the model is ready. Of course, changes of variable are often useful.
This model is much simpler when we work with the log-odds $\nu$ in favor of $+1$, defined as $\nu_i=\ln(\frac{p_i}{1-p_i})$.  Bayes Theorem causes a change in $p_i$ that translate to a simple additive process in $\nu_i$. That is, if the neighbor supports $+1$, $\nu_i$ changes to $\nu_i+\alpha$, where $\alpha=\ln(\frac{a}{1-a})$; if the neighbor supports $-1$, $\nu_i$ changes to $\nu_i-\alpha$. That is, the model is a simple additive biased random walk, with the bias dependent on the choice of the neighbors of each agent.

When the spatial structure is introduced, simulations have shown~\cite{martins08a,martins08b} that the emerging consensus is only local. Neighborhoods that support one idea will reinforce themselves and, with time, most of the agents become more and surer of their opinions, to the point they can be described as extremists. An extremist is defined as someone who is very close to be sure about one issue (very large $|\nu_i|$, corresponding to $p_i$ very close to certainty, 0 or 1). This happens even when all the agents had moderate opinions as initial conditions, unlike other models, where extremists have to be artificially introduced from the beginning. One should notice that the underlying continuous opinion allows us to speak of strength of opinions, unlike typical discrete models and, as such, at first, it is not so clear how CODA relates to those models.

When analyzed using the framework, it is clear how one can generalize CODA model to different scenarios.  Per example, by modeling a situation where $\alpha\neq\beta$ and $\beta$ is a function of time, it was possible to obtain a diffusive process from the CODA model where the diffusion slows down with time until it freezes~\cite{martinspereira08a}, with clear applications in the spread of new ideas or products. By modeling the influence of Nature as a bias in the social process of Science, CODA also proved useful to improve the understanding of how scientific knowledge might change \cite{martins10a}.

As an extension of the model, we can assume that the likelihoods depend not only on the opinion of the neighbor, but also on the agent's own observed choice. This is equivalent to introducing in the agent some awareness that its neighbor's choices might be dependent not only on the best choices, but could also be a reflection of its own influence upon that neighbor. For calculation purposes, assume, without lack of generality, that the first agent choice is $s_i=+1$. That is, the likelihood $P(s_j=+1|x=1)$ is replaced by two different probabilities 
\[
a=P(s_j=+1|x=+1,s_i=+1)
\]
\begin{equation}
\neq P(s_j=+1|x=+1,s_i=-1)=c
\end{equation}
and $P(s_j=-1|x=-1)$ is replaced by 
\[
b=P(s_j=-1|x=-1,s_i=-1)
\]
\begin{equation}
\neq P(s_j=-1|x=-1,s_i=+1)=d.
\end{equation}

Solving the Bayes Theorem and calculating the log-odds of the opinion, if the neighbor agrees ($s_i=+1$), we have 
\begin{equation}\label{eq:agreement}
\nu(t+1)=\nu(t)+\ln\left(\frac{a}{1-d}\right),
\end{equation}
and, if there is disagreement,
\begin{equation}\label{eq:disagreement}
\nu(t+1)=\nu(t)+\ln\left(\frac{1-a}{d}\right).
\end{equation}
The steps will only be equal in modulus, aside different signs, if $a=d$. This corresponds to the situation where both $x=+1$ and $x=-1$ are equally strong in influencing the agents and the agent $i$ choice is considered irrelevant for the choice of its neighbor $j$. On the other hand, if the agent $i$ considers that, when $s_i=+1$, it is more likely that a neighbor will choose $s_j=+1$, than we must have $a>d$. In this case, the steps will not have the same value and disagreement will have a more important impact than agreement.

The case where $a\rightarrow 1$ is interesting. If $a=1$ exactly, agent $i$ expects that, whenever it chooses $a$ and $x=+1$ is actually the best choice, the neighbor $j$ will also choose $x=+1$ with certainty. That means that an observation of $s_j=+1$ carries no new information,  while $s_j=-1$ would actually prove that $x=+1$ can not be the better choice. What happens is that, when $a=1$, the problem is no longer probabilistic, but one of Classical Logic. And as soon as the agent observes both decisions on its neighbors, it is faced with an unsolvable contradiction, unless $a$ is not exactly 1, but only close to. That is, we can work with the limit $a\rightarrow 1$, but $a$ should actually never be exactly 1.

Calculating the limits of the steps in Equations~\ref{eq:agreement} and ~\ref{eq:disagreement}, we have, for the agreement case,
\begin{equation}
\lim_{a\rightarrow 1}\left(\ln\left(\frac{a}{1-d}\right)   \right)= L,
\end{equation}
where $L$ is finite and non-zero. For disagreement, on the other hand, we have 
\begin{equation}
\lim_{a\rightarrow 1}\left(\ln\left(\frac{1-a}{d}\right)  \right)\rightarrow -\infty.
\end{equation}

As we go to the limit, agreement will tend to cause a negligible change to the value of $\nu$, when compared with the change caused by disagreement. If all agents start with reasonably moderate opinions, so that, whenever they find disagreement, their choices will flip, the system, in the $a\rightarrow 1$ case is a simple one. Whenever the neighbor agrees, the first agent will not update its opinion (or update very little, if $a$ is not exactly 1). When the neighbor disagrees, the first agent will change its observed opinion to that of the neighbor. In other words, when agent $i$ observes agent $j$ choice, it always end with the same choice as $j$. In the limit, we obtain the traditional voter model~\cite{cliffordsudbury73,holleyliggett75}. 

That is, we basically have a dynamics where the agent only updates its mind when there is disagreement. This same update dynamics is observed in other discrete models, as per example, for Sznajd interactions~\cite{sznajd00,stauffer03a,sznajd05}. In Sznajd model, it takes two agreeing agents to convince all other neighbors. Basically, it works the same way as the voter model, except for the description of when an interaction happens. Since the Bayesian framework is only applied here to the opinion update and not to the rules of interaction, we have the same case as we had in the voter model. Other features, such as contrarians~\cite{galam04}, are also easily introduced by a simple change in the likelihood. If an agent considers its neighbor more likely to be wrong than correct, the agent opinion will change away from that of the neighbor, hence, a contrarian \cite{martinskuba09a}. 

Finally, the models of hierarchical voting~\cite{galam2003a,galam06b}, where the decision of each level is obtained from the majority of the voters, except when there is a tie, can also be easily translated into CODA Bayesian language using the same strategy as in the voter model. That is, agreement with the majority means no reinforcing of previous opinion, while disagreement leads to an observable change. If there is a slightly different likelihood in favor of one theory, when there is a tie, that theory will tend to be picked up. The same effect could also happen due to small differences in the probabilistic continuous views of the individuals in the tie groups. Interestingly, the translation of the problem into CODA formalism suggests natural extensions of the model, where the final opinion might depend also on the continuous probability each agent assigns to each proposition.

\section{Agreement versus disagreement}\label{sec:selfinfluence}

Equations \ref{eq:agreement} and \ref{eq:disagreement} mean that different values are added when there is agreement and when there is agreement. Since only the sign of $\nu$ is important, if all $\nu$s are multiplied by the same arbitrary constant, the system remains the same. That means that we can choose, without lack of generality, the value of the additive constant to be $1$ when there is disagreement and a proportional step size $S$ when there is agreement. For the original CODA model, $S=1$. And $S<1$ corresponds to the case where an agent thinks his neighbor is influenced by his own opinions (this is not about cause and effect, just a probabilistic assessment that they are more likely to agree). The limit discussed in the previous Subsection will happen when $S\rightarrow 0$. Since $S$ measures the relative importance between disagreement and agreement, we can, in principle, study also the case where $S>1$. This choice, $S>1$, corresponds to the agents assuming their neighbors tend to disagree with them. This might not be a very realistic assumption for many applications, but we will explore the whole range of values for $S$, for completeness sake and in order to understand the problem better.

First, as pointed out in the previous Section if the system will run so that each agent will interact with others $T$ times in average, if $S$ is significantly smaller than $1/T$, successive agreements will add less than 1. Since a disagreement adds 1 in the opposite direction, this is the region where the model will be indistinguishable from a Voter model. That means that  interesting and new effects, different from voter model results, should happen when $S>>1/T$.

\begin{figure}[ph]
\centering
\begin{tabular}{cc}
\psfig{file=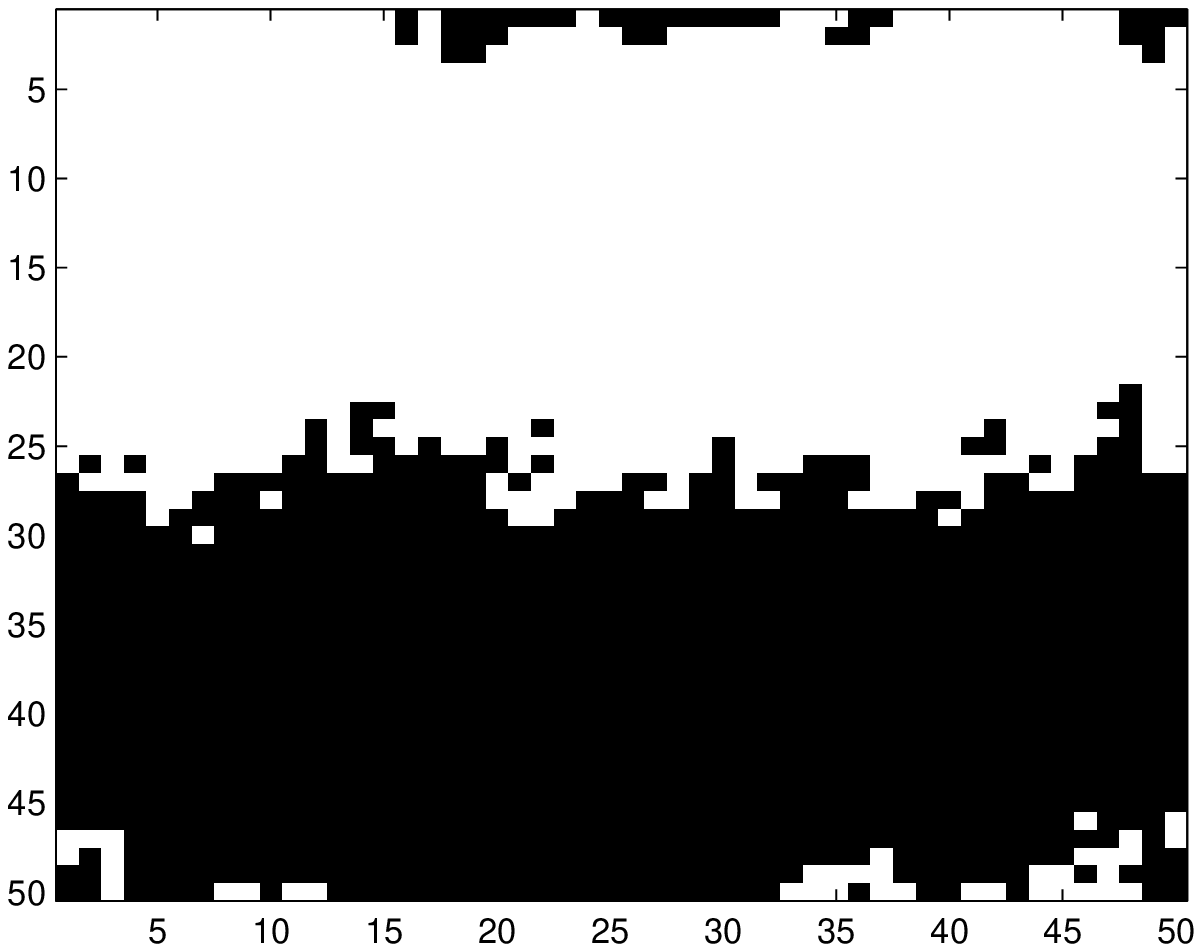,width=6.0cm} & 
\psfig{file=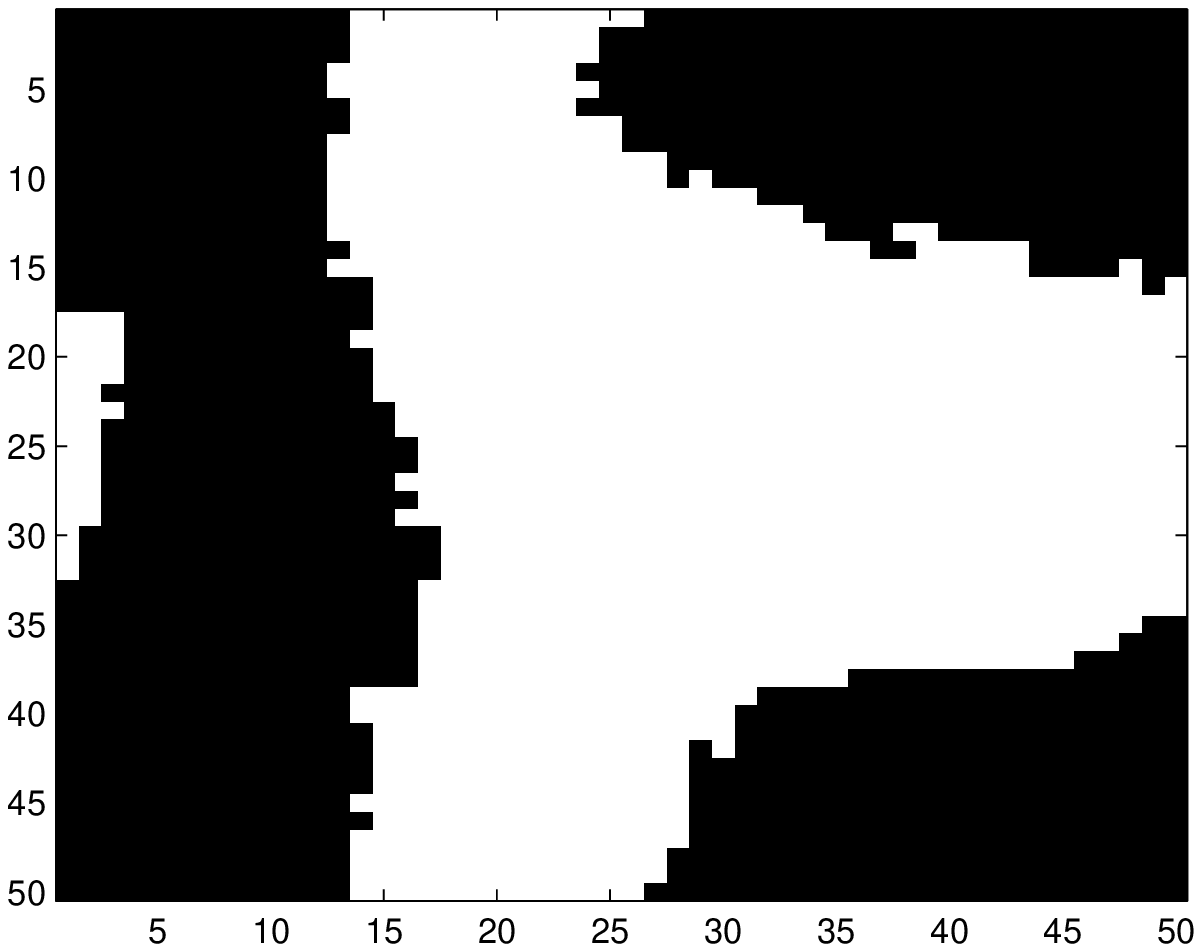,width=6.0cm} \\
\psfig{file=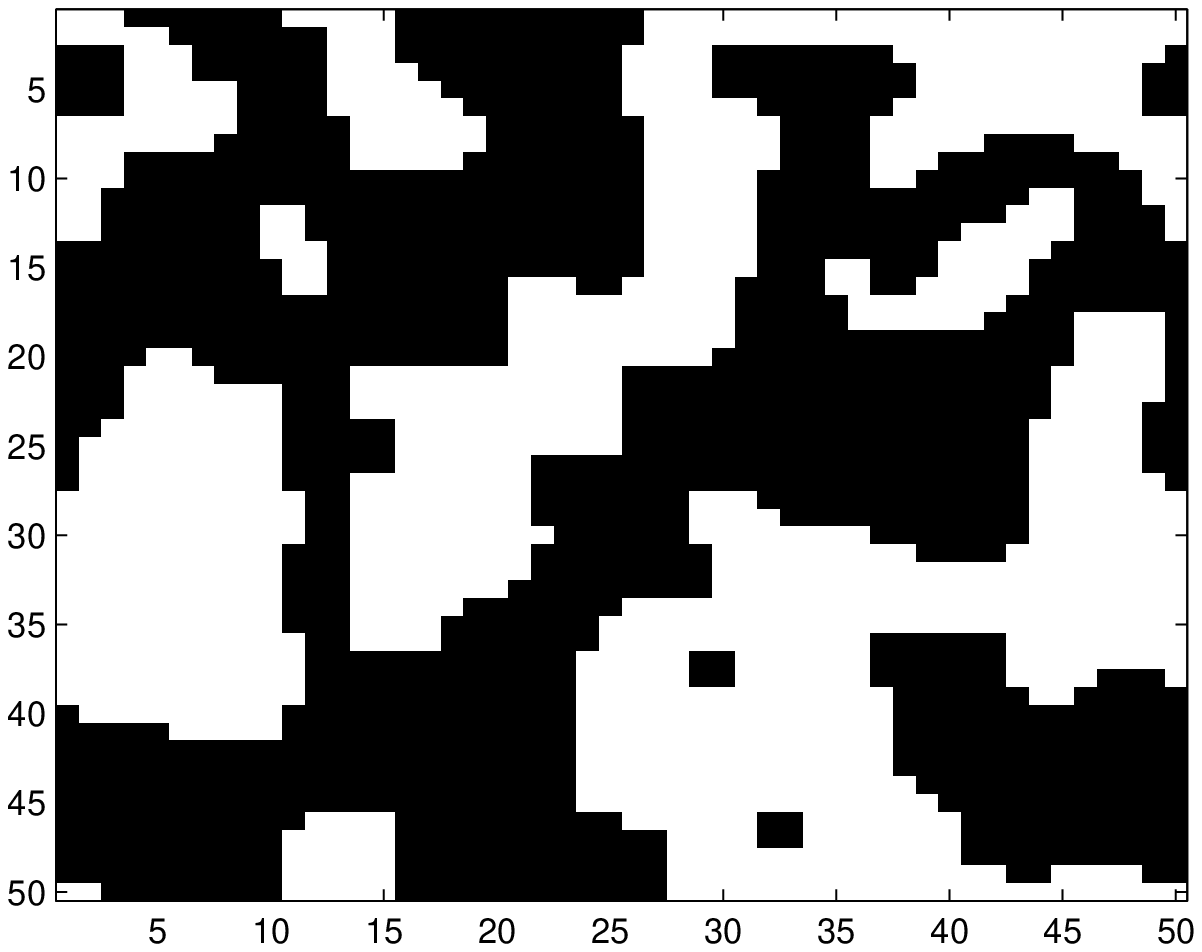,width=6.0cm=} &
\psfig{file=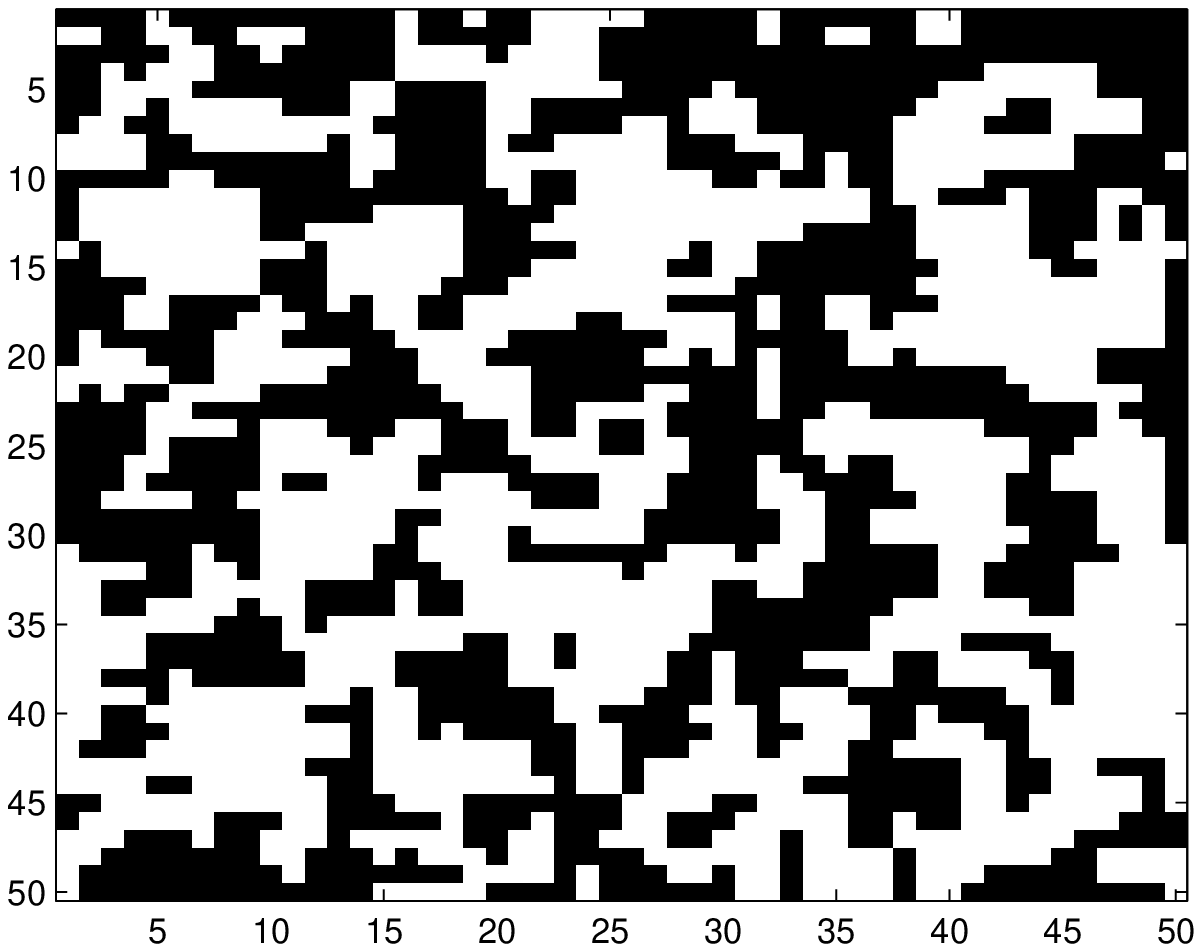,width=6.0cm}
\end{tabular}
\vspace*{8pt}
\caption{Typical configurations of choices for the middle of runs for different values of $S$ in a bi-dimensional square lattice with $n=50^2$ agents. Top lattices correspond to $S=0.01$ and $S=0.1$ and the lower ones to $S=1$ and $S=10$.}\label{fig:finalstates}
\end{figure}

The problem was implemented, for testing, in a square bi-dimensional lattice with periodic boundary conditions and first neighbors interactions. Several tests for different values of $S$ show that, unless the influence of agreement is small, no consensus emerges in the long run. Figure \ref{fig:finalstates} shows snapshots of simulations of the system for different values of $S$, from $S=0.01$ to $S=10$. What we see is that as agreement becomes weaker, larger areas with the same opinion appear. Consensus is still hard to achieve, but the increase in size points to the limit where only one area will dominate the full system. The tendency is reversed for larger $S$. $S=1$ corresponds to the original CODA model, with clear regions adopting each choice, but still large interfaces between them. For $S=10$, what we see is just a very weak tendency for similar opinions to be together. This happens because as soon as one neighbor agrees with an agent, the tendency for the opinion of each of them is to be reinforced, even when surrounded by disagreers. In the simulation, each agent had 4 neighbors, meaning there would be $1/4$ of chance of reinforcing, adding 10 and $3/4$ of chance of weakening the opinion by just 1. As long as there is one neighbor that agrees, each agent tends to keep his opinion. The global effect is that no clear regions appear, with most agents living in a mixed neighborhood.

The snapshot for the $S=0.01$ case, in the upper left of Figure \ref{fig:finalstates}, has clues on how the system makes the transition from reinforcement of opinions inside each domain to be able to eventually achieve consensus. Notice that the space is basically divided in two zones separated by straight lines with large fluctuations at the borders. Since $S=0.01$, any agent on the interface has a tendency to change its choice, since the influence of disagreement is much stronger. This causes both opinions to invade the opposing view. However, the deeper an invasion goes, the more likely it will encounter agents who had just had their opinions reinforced until then. That means that it is more likely that the agents in the invasion will be convinced back, since their opinions are new, than they will remain long enough to compensate for the previous reinforcement.  

\begin{figure}
\centering
\begin{tabular}{c}
 \includegraphics[width=0.75\textwidth]{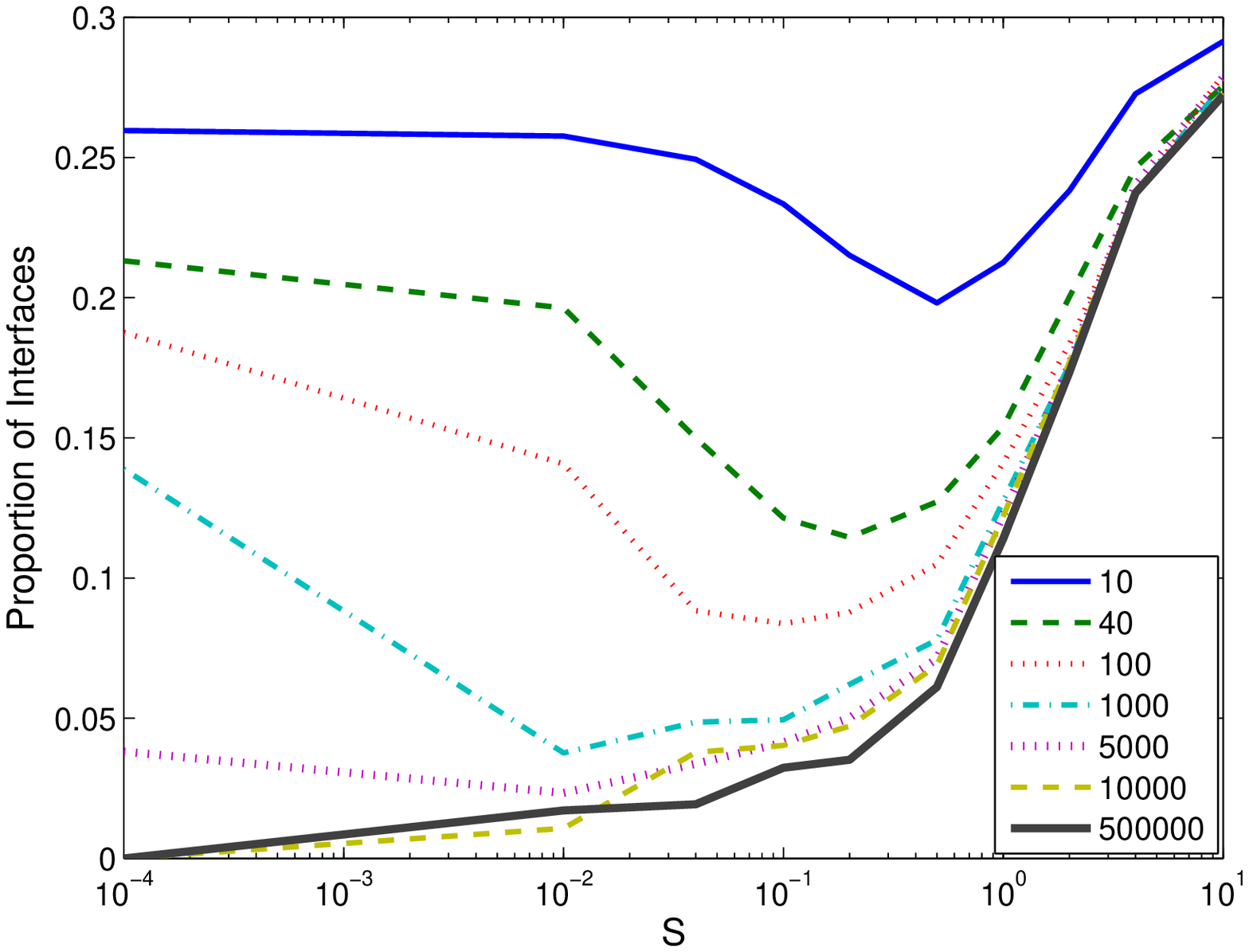} \\
\includegraphics[width=0.75\textwidth]{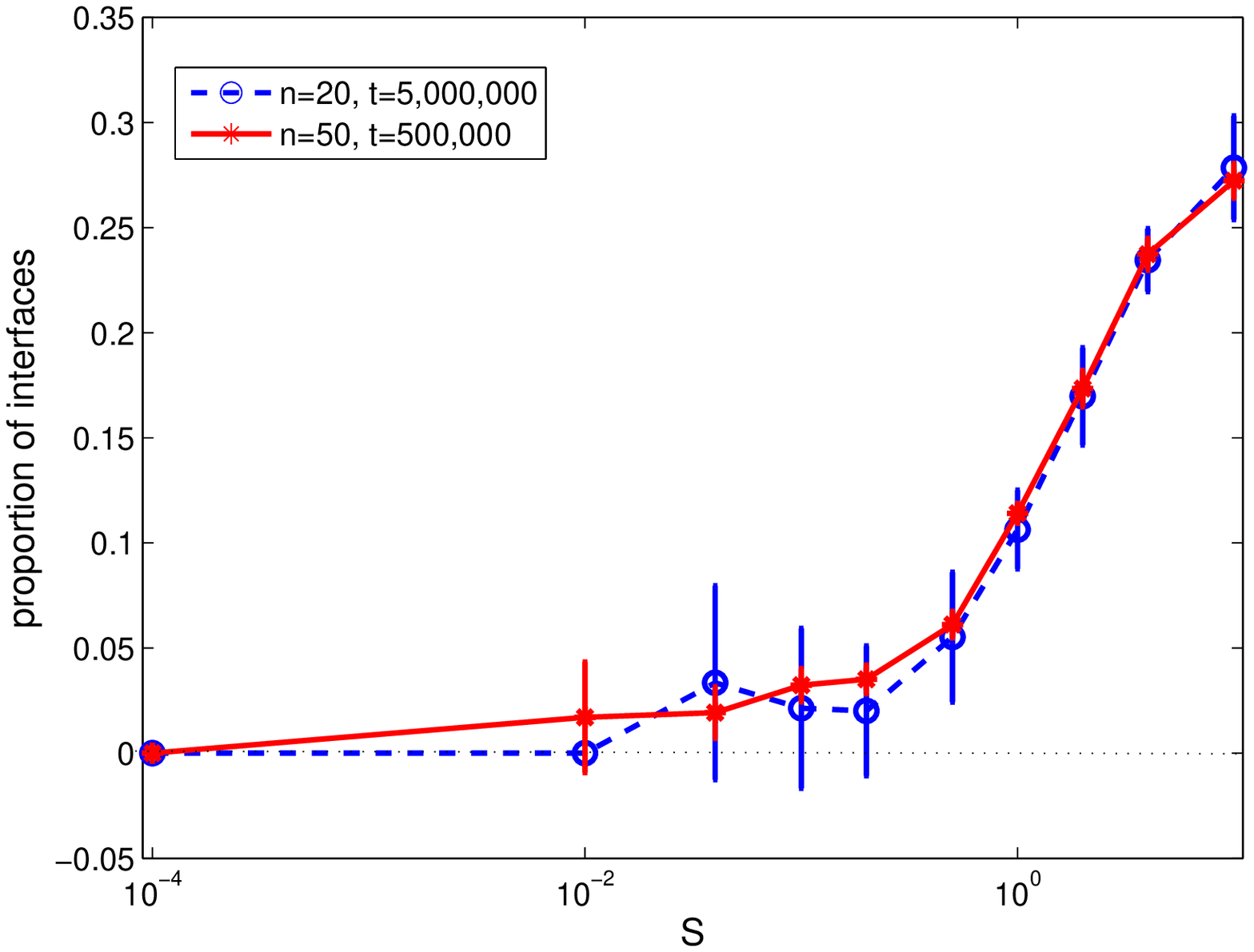}
\end{tabular}
\caption{Average proportion of neighborhood links that correspond to interfaces (different choices), as a function of $S$. {\it Upper Panel}: Evolution of the interface proportion curves for different average number of interactions per agent $t$ for a square network with $n=50^2$ agents. {\it Lower Panel}:  Evolution of the interface for a network with $n=20^2$ after 5,000,000 of average interactions per agent and $n=50^2$ after 500,000 interactions, both shown with error bars representing the standard deviation of the results of different realizations of the problem.}\label{fig:interfaces}
\end{figure}

Figure \ref{fig:interfaces} shows how the interfaces between the two competing choices are distributed for different values of $S$ and also how they evolve in time. Each point correspond to the average (over 20 realizations) proportion of links where the agents have opposing views, relative to the total number of links. The top panel shows different numbers $t$ of average interactions per agent, all for a network with $n=50^2$ agents, while the bottom panel shows the results for two different network sizes, $n=20^2$ and $n=50^2$ after $t=5,000,000$ and $t=500,000$ average interactions per agent respectively with error bars indicating the standard deviation of the observed values over the different realizations of the system.

We observe that, for small number of interactions, there is a valley in the proportion of interfaces at intermediary values of $S$. The increase in the proportion of interfaces happens, for very large $S$, because  as we have seen, the systems freezes fast in a mixed condition. For small $S$, a different condition is observed. There is a clear tendency for the system to organize in different regions and that should lead to a small proportion of interfaces. But, as we have discussed, invasions of one region into the other are very common. This actually prevents the large regions from forming soon and the proportion of interfaces from going down fast. However, as $T$ becomes larger and larger, that proportions starts decreasing steadily. This corresponds to the transition to two well defined regions and, eventually, complete agreement, when the proportion of interfaces is trivially zero. 

We also see that for smaller values of $S$ the invasions can eventually take the system to agreement. This happened, for the $n=50^2$ lattice  after $T=500,000$ average interactions per agent in all realizations when $S=0.01$ and some, but not all, when $S=0.1$. For the smaller network ($n=20^2$), running for a longer time ($T=5,000,000$) realizations ending in full consensus were observed even when $S=0.4$. It is important to notice that in Figure \ref{fig:finalstates}, the domains for $S$ smaller than one but not too small were very large. Fluctuations in a small network can spread through all the agents more easily. The reason why consensus emerged in those cases is due to finite size effects. 
 
That is, as agreement becomes a weaker force, a number of interesting features happen. If it is small enough, the system might be taken to consensus, even though this consensus can require so many interactions that the scale of time needed for that is much larger than that in any real social system. For small, but not so small values of $S$, we observe a tendency to the existence of larger groups. While smaller $S$ will mean that movements of reinforcement are smaller and, therefore, inside the regions, opinions will be strong but not as strong as in the CODA model, the smaller proportion of interfaces mean that more agents are shielded from interacting with different opinions. The more extreme opinions are a less strong, as agreement is weaker. But with smaller proportion of interfaces, more agents will have their opinions changing towards their own choice, meaning that less strong but still extreme opinion will be shared by more agents. Surprisingly, by making the influence of those you agree with weaker, the system is lead to a state where more agents share the extreme point of views.

\section{Conclusions}

We saw that Bayesian rules can provide a theoretical basis to model the change in the opinion of agents in both a more realistic and more flexible way, probably a little closer to how real people think. The introduction of these ideas in a Ising-like scenario, where only binary choices can be observed, had made it possible the modeling of the emergence of extremism  in the context of the CODA model. Here, we have seen how several traditional models of the literature can be obtained as a limit case of a variation of the CODA model. 

We have also seen in the agreement versus disagreement model that, by considering one's own effect on one's neighbor, interesting and unexpected effects are observed. While accounting for one's own effect does make the effect of observing an agreement weaker on one's opinion than observing a disagreement, this weakening has two main effects. One is that opinions inside a domain are extreme, but less so, as they are trivially reinforced by a smaller value. But the system as a whole gets organized in a way that less agents observe any disagreement. That means that more agents will actually have those strong opinions.

\section{Acknowledgments}

The author would like to thank both Funda\c{c}\~ao de Amparo \`a  Pesquisa do Estado de S\~aoPaulo (FAPESP),  under grant  2009/08186-0,  for the support to this work.

\bibliographystyle{unsrt}
\bibliography{biblio}

\end{document}